\documentclass[pra,bibnotes,twocolumn]{revtex4}
\usepackage{graphicx}
\begin{document}
	\draft
	\def\ds{\displaystyle}
	\title{ Diffraction-free beam propagation at the exceptional point of non-Hermitian Glauber Fock lattices
	}
	\author{Cem Yuce$^1$}
	\email{cyuce@eskisehir.edu.tr}
	\author{Hamidreza Ramezani$^2$ }
	\address{$^1$Department of Physics, Faculty of Science, Eskisehir Technical University, Turkey \\
		$^2$Department of Physics and Astronomy, University of Texas Rio Grande Valley, Edinburg, TX 78539, USA}
	
	\date{\today}
	\begin{abstract}
		We construct localized beams that are located at the edge of a non-Hermitian Glauber Fock (NGF) lattice made of coupled waveguides and can propagate for a long distance without almost no diffraction. Specifically, we calculate the closed-form of the eigenstates at the exceptional point of a semi-infinite NGF lattice composed of waveguides which are coupled in a unidirectional manner. We use the closed-form solution to construct the non-diffracting beams in finite NGF lattices. We provide a numerical approach to find other lattices that are capable of supporting non-diffracting beams at an exceptional point.
	\end{abstract}
	\maketitle

	\section{Introduction}
	The physics of photon propagation in discrete lattices is very rich and has been extensively studied \cite{dcn} which resulted in the introduction of synthetic lattices capable of supporting non-diffracting wave propagation. This includes lattices supporting Airy beams \cite{aridc} 
	or flat bands in Hermitian lattices \cite{8,9} and non-Hermitian parity-time symmetric lattices \cite{rmf,konotop,alexfb}. Specifically, in parity-time symmetric lattices, the flat band can occur at the exceptional point (EP) which allows for non-diffracting beam propagation. Exceptional points are topological singularities in non-Hermitian Hamiltonians \cite{ep000k}. At an $n$th order exceptional point, $n$ eigenvalues and the corresponding eigenstates of a non-Hermitian Hamiltonian coalesce. Exceptional points are ubiquitous in optics and have interesting transport feature which its properties manifested in manipulating light propagation such as unidirectional invisibility \cite{ep1,ep2}, unidirectional lasing \cite{ep3}, lasing and anti-lasing in a cavity \cite{ep4} and enhanced optical sensitivity \cite{ep5}.

	One special system where discrete diffraction can be studied is the semi-infinite and asymmetric Glauber Fock lattice.  Hermitian Glauber Fock lattice has recently been implemented and demonstrated in optical lattices \cite{glauber0,glauber3,glauber4}. A Glauber Fock photonic lattice is composed of an array of evanescently coupled waveguides with a square-root distribution of the coupling between adjacent waveguides \cite{glauber0}. The first experimental realization with direct observation of the classical analog of Fock state displacements was reported in \cite{glauber3}. The Glauber-Fock photonic lattice is interesting in the sense that every excited waveguide represents a Fock state and an infinitely long lattice admits an exact analytical solution.

	All eigenvalues and the corresponding eigenstates coalesce at the $N$th order exceptional point in a finite lattice with $N$ sites. In this case, there is a unique exceptional eigenstate in the system. In this paper, we show that this statement is not true in a semi-infinite lattice by getting the most general form of exact analytical solution for a non-Hermitian Glauber-Fock lattice. We show that there is a continuous family of eigenstates at the infinite order exceptional point. In practice, every lattice has a finite number of lattice sites and thus one might think that the continuous family of eigenstates remains a mathematical curiosity. However, here we show that the continuous family of eigenstates \cite{CYHR20} can be used to construct non-diffracting waves at the exceptional point in a finite Glauber-Fock photonic lattice. We remind the reader that this is similar to self-accelerating waves, which is non-integrable and physically impossible to be realized. However, truncated self-accelerating waves have been considered as non-diffracting waves up to a large distance in an experiment.
	
	\section{Model}
	Consider a nonreciprocal 1D lattice with $N$ lattice sites described by the non-Hermitian Hamiltonian
	\begin{eqnarray}\label{hjd?akl1}
H\psi_n=\gamma_{n+}~ \psi_{n+1}+\gamma_{n-}~ \psi_{n-1}
\end{eqnarray}
where $\ds{\gamma_{n+}}$ and $\ds{\gamma_{n-}}$ describe the site-dependent forward and backward hopping amplitudes, respectively and $\psi_n$ is the complex field amplitude. The system has an $N$th order exceptional point at either $\ds{\gamma_{n+}=0}$ or $\ds{\gamma_{n-}=0}$. In this case, all eigenstates coalesce to a unique exceptional state with zero eigenvalue. It is obvious that for $N\to \infty$ the EP turns to be an infinite order EP. At this point we are interested in finding the exact form of such infinite order EP. From mathematical point of view, recently, we have shown that the eigenstates of semi-infinite lattices may not coalesce at a specific value \cite{CYHR20}. Here, first, we will explore this fact in an analytically solvable model. The model we consider here is a variant of semi-infinite Glauber-Fock photonic lattice, where the hopping amplitude increases with the square root of the site number, $\ds{\gamma_{n+}=\gamma_{+} \sqrt{n+1}}$ and $\ds{\gamma_{n-}=\gamma_{-}\sqrt{n}}$. Here $\ds{\gamma_{+}}$ and $\ds{\gamma_{-}}$ are constants describing site-independent forward and backward hopping amplitudes, respectively. Therefore the equation satisfied by the complex field amplitude $\psi_n$ at the $n$-th waveguide is given by 
	\begin{eqnarray}\label{ham23}
-i\partial_z\psi_n+\gamma_+ \sqrt{n+1}~\psi_{n+1}+\gamma_{-}\sqrt{n} ~ \psi_{n-1}=0
\end{eqnarray}
	where $\ds{z}$ is the normalized propagation distance, $\ds{n=0,1,2,...,N-1}$. Note that the system is non-Hermitian when $\ds{\gamma_+\neq\gamma_-}$. 
	
	Let us look for stationary solutions of the form $\ds{\psi_n(z)=  e^{-iE z}  \psi_n (0)}$. Note that for finite $\ds{N}$, all eigenstates coalesce with zero energy eigenvalue. If $\ds{\gamma_{-}=0}$ and $\ds{\gamma_{+}\neq0}$, the exceptional state is well localized at the right edge and its matrix representation is given by $\ds{\psi_n(0)=\{ 0,0,...,1\}^{T}  }$, where $T$ denotes the transpose. On the other hand, if $\ds{\gamma_{+}=0}$ and $\ds{\gamma_{-}\neq0}$, the exceptional state is well localized at the left edge and is given by $\ds{\psi_n(0)=\{ 1,0,...,0\}^{T}  }$. We stress that this localization character of the exceptional eigenstate is valid independent of $N$. However, this is not the case if $N$ is $\infty$, i. e., for the semi-infinite lattice.  If $\ds{\gamma_{+}=0}$, such a coalescing state is not available since there is no right edge in the semi-infinite lattice. Surprisingly, in the case of semi-infinite lattice with $\ds{\gamma_{+}=0}$, we get a continuous family of eigenstates instead of a unique exceptional state even if the system is at the exceptional point. To obtain the most general form of the exact analytical solution, we write the state vector as $\ds{| \Phi  >=\sum_{n=0}^{\infty} \psi_n (z) |n>}$, where the Fock state $\ds{|n>}$ corresponds to a situation when only the waveguide with number $n$ is excited \cite{glauber3,glauber0}. Substituting this solution into the equation (\ref{ham23}) yields the Schrodinger-like equation $\ds{\mathcal{H}\Phi=i\frac{\partial\Phi}{{\partial} z}}$, where time is replaced by the propagation distance. The corresponding Hamiltonian reads $\ds{
		\mathcal{H}=  \gamma_+ ~ \hat{a}^{\dagger}+~\gamma_{-}~\hat{a }
	}$. Here $\ds{a^{\dagger}}$ and $\ds{a}$ are the well-known bosonic creation and annihilation operators satisfying $\ds{\hat{a}^{\dagger}|n>=\sqrt{n+1}|n+1>}$ and $\ds{\hat{a}|n>=\sqrt{n}|n-1>}$, respectively. We can transform this Hamiltonian using $\ds{\hat{a}=\frac{q+ip}{\sqrt{2}}}$ and $\ds{\hat{a}^{\dagger}=\frac{q-ip}{\sqrt{2}}}$, where $\ds{q}$ and $\ds{p}$ are the normalized position and momentum operators, respectively. Then the Hamiltonian can be rewritten in the following form $\ds{\mathcal{H}=  \frac{i(\gamma_{-}-\gamma_+ ) }{  \sqrt{2} }p+\frac{\gamma_{-}+\gamma_+  }{  \sqrt{2} } q     }$. Notice that we assume $\ds{\gamma_{-}\neq\gamma_+ }$, which is the condition for non-Hermiticity. Let us substitute $\ds{p=-i \partial_q}$ and solve the corresponding Schr\"{o}dinger-like equation. The resulting equation is of first order and admits an exact analytic solution. If $\ds{\gamma_+}$ and $\ds{\gamma_-}$ are z-dependent, then the most general form of the solution is given by	
	\begin{equation}\label{plkjdqats}
	\Phi(q,z)=e^{   -\int_0^qS(Z+iq-is) s ds             }~~F( Z+i  q)
	\end{equation}
where $\ds{S(Z)= \frac{\gamma_{-}+\gamma_+  }{  \gamma_{-}-\gamma_+}  }$, where $\ds{Z= \int_0^z\frac{    \gamma_{-}-\gamma_+    }{  \sqrt{2}  }   dz }$ and $\ds{F(q)}$ is an arbitrary continuous function. If $\ds{\gamma_+}$ and $\ds{\gamma_-}$ are constants, then it is reduced to
	\begin{equation}\label{plkjdqats}
	\Phi(q,z)=e^{-\frac{\gamma_{-}+\gamma_+ }{2 (\gamma_{-}-\gamma_+ )} q^2 }f( z+ \frac{ \sqrt{2}i}{\gamma_{-}-\gamma_+  } q)
	\end{equation}
Note that we made the replacement $\ds{ F( \frac{   \gamma_{-}-\gamma_+  }{    \sqrt{2}  }   z+iq) \rightarrow    f( z+ \frac{ \sqrt{2}i}{\gamma_{-}-\gamma_+  }  q)   }$ in the last step. We stress that we do not require to have $\ds{\Phi(q  \rightarrow \mp    \infty)=0  } $ while we require to satisfy $\ds{\psi_n(n  \rightarrow  \infty)=0  } $. The arbitrary function $\ds{f(q)}$ can be determined from the initial condition at $z=0$. Equation (\ref{plkjdqats}) allows us to find the time evolution of any initial state. In $q$-space, the wave packet translates with a constant speed. In the original space, this may imply growing or decaying diffracting solutions depending on the form of $f(q)$. 

Considering the solution in q-space (\ref{plkjdqats}), let us now find the form of the wave packet in the original space, which is given by $\ds{\psi_n(z)=<n|\Phi>}$
	\begin{equation}\label{uf501owkjts}
	\psi_n(z)=\int_{-\infty}^{\infty}\frac{\pi^{-1/4} }{\sqrt{2^nn!}}   e^{-q^2/2} H_n(q) ~\Phi(q,z) ~dq 
	\end{equation}
	where  $H_n(q)$ is the nth order Hermite polynomials. This is the most general solution of our original system and one can analytically study not only the eigenstates but also time evolution of any given initial wave packet for the semi-infinite non-Hermitian lattice at the exceptional point. 
	
As a special case, consider $\ds{\gamma_{-}=-\gamma_+}$. In this case, the transformed Hamiltonian is reduced to $\ds{\mathcal{H}=    \sqrt{2} i\gamma_-  p}$, which is an anti-Hermitian Hamiltonian $\ds{  \mathcal{H}=-\mathcal{H}^{\dagger}}$. In this case the solution (\ref{plkjdqats}) becomes $\ds{   \Phi= f( z+ \frac{i}{ \sqrt{2}\gamma_{-}   }  q)  }    $. We get a specific zero energy stationary solution if we choose $f=1$. One can choose different from of function $f$ such as $f(q)=e^{- q^2/4}$ and $f(q)=e^{ q^2/4}$, which are not stationary solutions (they are decaying and growing solutions). 
	
Next, let's calculate the eigenstates of our original Hamiltonian. Let the arbitrary function $f$ in Eq. (\ref{plkjdqats}) to be
	\begin{equation}\label{w74afbats}
	f_E=\exp{        \left( -i E ( z+ \frac{ \sqrt{2}i}{(\gamma_{-}-\gamma_+)  }  q)    \right)       }
	\end{equation}
	where the constant $E$ are eigenvalues. Below, we show that $E$ does not take discrete values but continuous values. One can directly substitute them into the integral (\ref{uf501owkjts}) to get the continuous family of eigenstates with energy $E$. One can see that as the contrast between $\ds{\gamma_+}$ and $\ds{\gamma_-}$ increases, the localization length of the eigenstates get increased.  	
	\begin{figure}[t]
		\includegraphics[width=4cm]{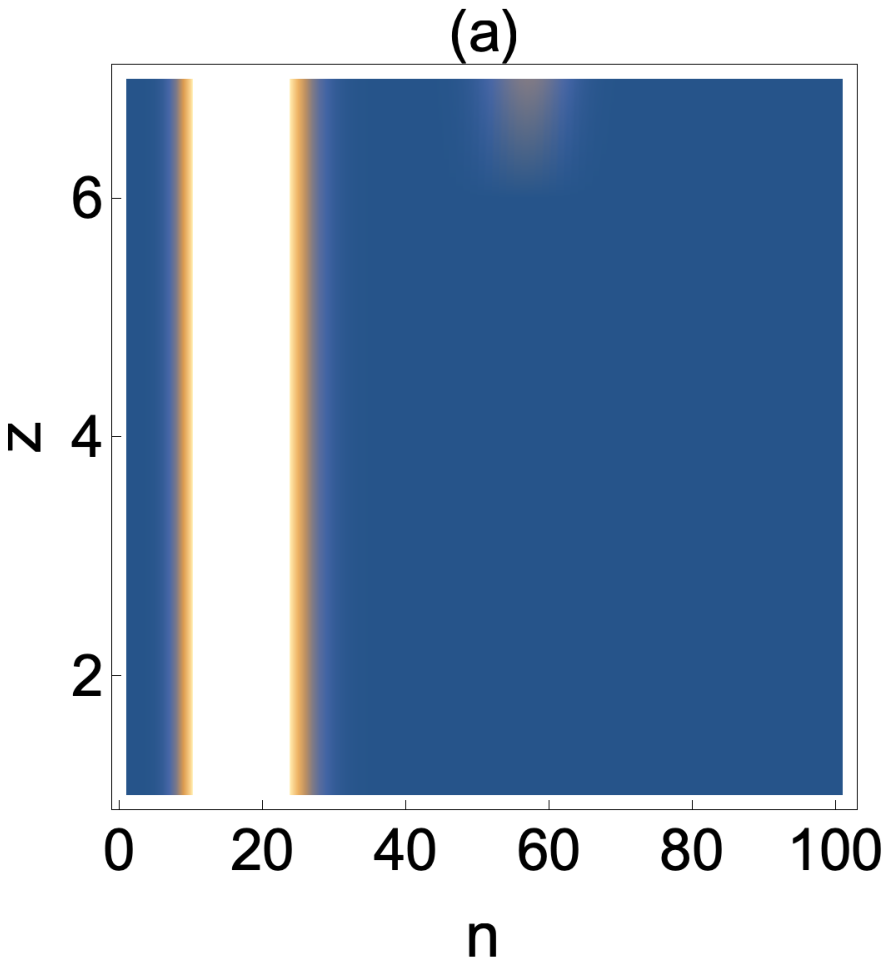}
		\includegraphics[width=4cm]{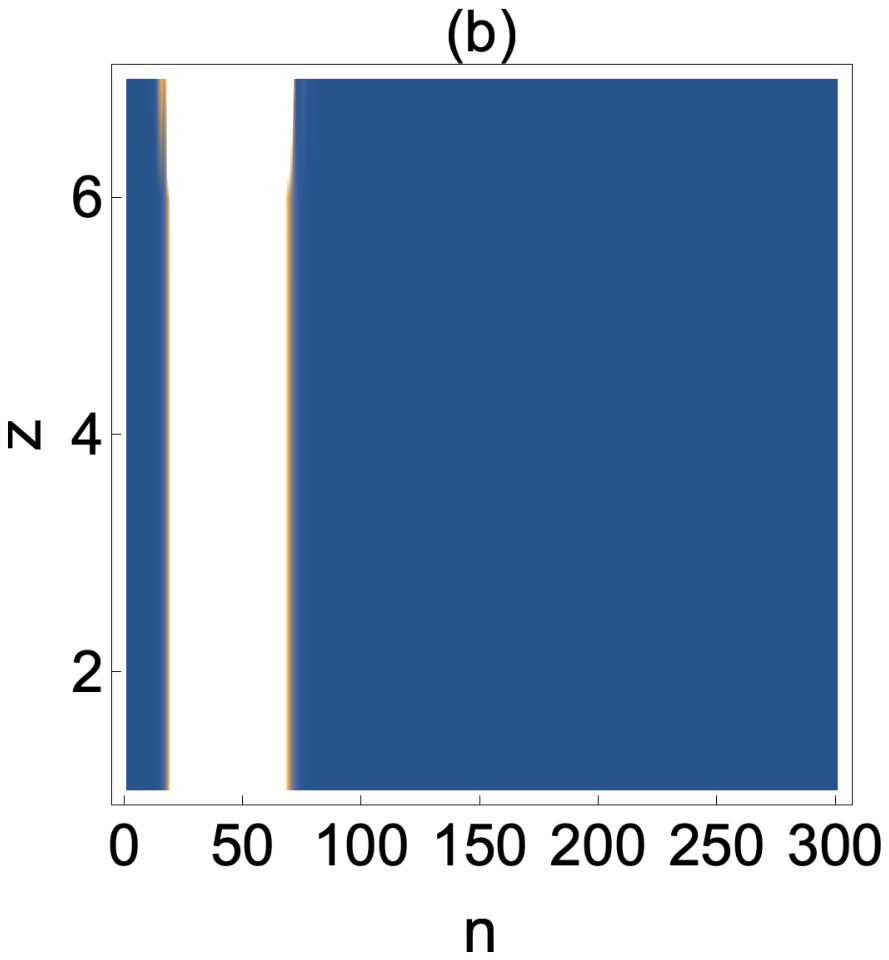}
		\caption{ The density plots for $N=100$ (a) and $N=300$ (b). $\ds{\alpha}$ is chosen a larger value in (b) to shift the center of the wave packet to the right. At the exceptional point $\ds{\gamma_+=0}$, we can construct non-diffracting wave packet. Note that the true exceptional eigenstate is the one where only the $n=0$th lattice site is excited.}
		\label{fig1}
	\end{figure}
	\begin{figure}[t]
		\includegraphics[width=4.2cm]{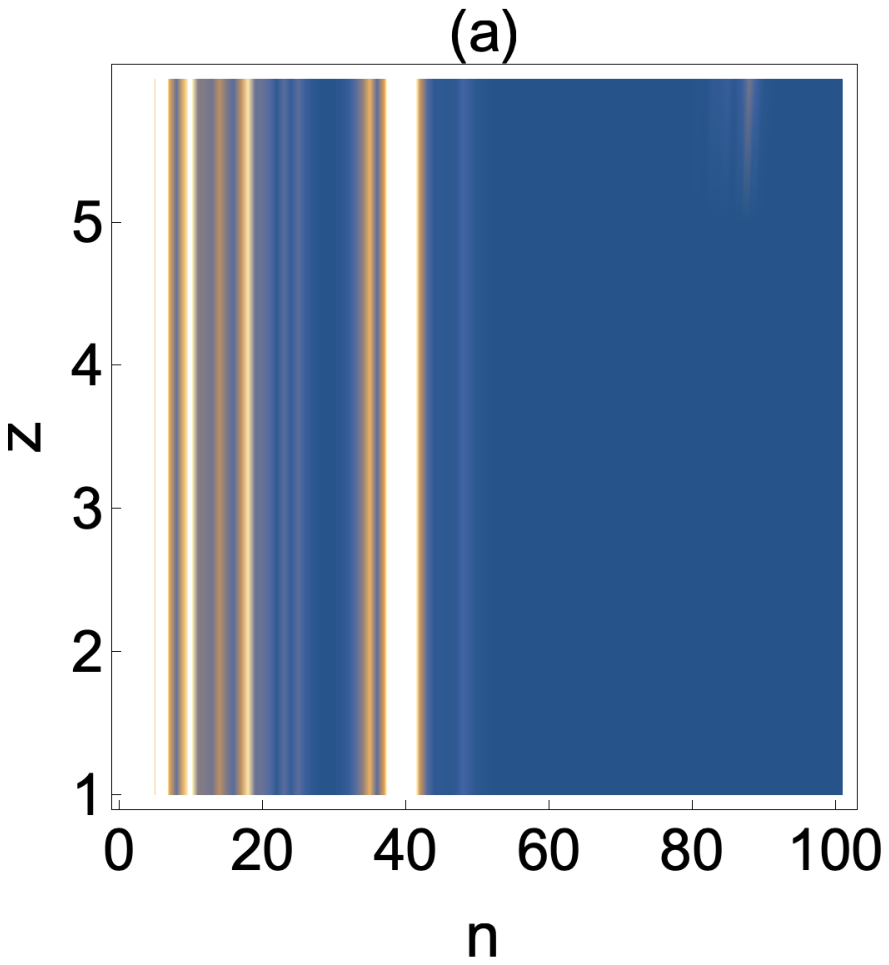}
		\includegraphics[width=4.2cm]{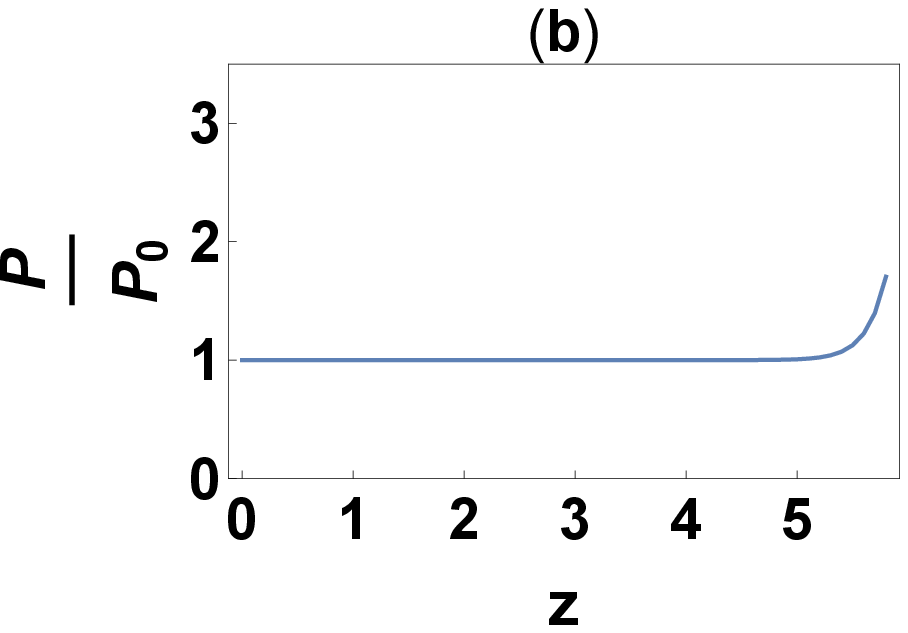}
		\caption{ The density plot for $N=100$ (a) and the ratio of total power to the initial one as a function of $z$ (b). The forward hopping amplitudes are set to to zero to obtain that the only eigenstate is the exceptional state well localized at the left edge. The backward hopping amplitudes are chosen randomly in the interval $\ds{\gamma_n\in[1,6]}$ and we take $\psi_{101}(z=0)=10^{-8}$. At around $z=5$, the non-diffracting character is lost ant the total power starts to increase exponentially. }
		\label{fig2}
	\end{figure}
	\subsubsection{Continuous family of stationary solutions at the exceptional point}
	Let us now explore stationary eigenstates at the exceptional point for the semi-infinite lattice. Suppose first that $\ds{\gamma_-=0}$ and $\ds{\gamma_+\neq0}$. In this case, the integral (\ref{uf501owkjts}) diverges since $\ds{\Phi=e^{\frac{q^2}{2}} f_E}$. This implies that no stationary solution is available. Note that the exceptional state is well localized at the right edge of a finite lattice. However, our lattice is semi-infinite and no exceptional state is possible to exist because the absence of edge where the exceptional state is localized. However, one can construct non-stationary solutions by choosing various form of $\ds{f(q)}$. 
	
	Suppose next that $\ds{\gamma_+=0}$ and $\ds{\gamma_-\neq0}$. For a finite lattice, the exceptional eigenstate is well localized at the left edge. Note that the well localized eigenstate $\ds{\psi_n=\{ 1,0,...\}^{T}  }$ at left-edge is also an eigenstate for the semi-infinite lattice. In fact, it is one of the zero energy solutions and we can construct infinite number of other eigenstates as a function of $E$. To see this, we evaluate the integral (\ref{uf501owkjts}) with (\ref{w74afbats}) and we find the well known coherent states, which satisfy the completeness condition but not the orthogonality condition
	\begin{equation}\label{otrpe9sts}
	\psi_n=e^{-|\alpha|^2/2}  \frac{\alpha^n}{\sqrt{n!}} 
	\end{equation}
	where $\ds{\alpha=\frac{ E }{\gamma_-} }$.  The constant $E$ is a continuous variable and hence this solution describes a continuous family of eigenstates. In other words, the unique character of the exceptional state is lost as $N\rightarrow \infty$. At $E=0$, $\ds{\psi_n=\delta_{n,0}}$, which implies that the zero energy eigenstate exactly matches the eigenstate well localized at the left edge. However, increasing $E$ continuously allows us to obtain a diverse family of stationary solutions at an exceptional point. Note that increasing $E$ shifts the center of the wave packet to the right. Let us require that $\ds{\psi(n\rightarrow\infty)=0}$. Then we obtain $\ds{|\alpha|\leq1}$ (the amplitudes in the expansion (\ref{otrpe9sts}) get decreased as $\ds{n}$ increases). To this end, we stress that $E$ can also take complex values. If $E$ is purely imaginary, then we can construct either decaying or growing solutions depending on the sign of $E$. Fortunately, the spatial form of the wave packet remains the same during the propagation. 
	
	Next let us discuss the practical application of the above solution since in a real experiment a lattice has a finite number of sites. Before discussing this point, we remind the reader about the self-accelerating waves, which is a non-integrable mathematical solution and hence physically impossible to be realized. However, it was shown that such solutions can still be used as non-diffracting waves up to a large distance if they were truncated \cite{saw1,saw2}. In a similar fashion, we claim that our solution can also be used to construct non-diffracting beams. In other words, truncating our solution allows us to construct almost stationary wave packets (or non-diffracting waves) in an experimental setup. To get practical non-diffracting waves at the exceptional point, let us truncate the solution (\ref{otrpe9sts}). This truncation woks if $N$ is large enough such that $\ds{   \frac{\alpha^N}{\sqrt{N!}}<<1 }$. In this case, the non-truncated terms are negligibly small and only contribute to the system in a very large time. Note that this truncated solution is not an exact analytical solution for a finite lattice with $N$ lattice sites as all eigenstates coalesce at the exceptional point. 
	
	Let us perform a numerical approach to support our idea. Suppose $\gamma_-=\sqrt{2}$, $\gamma_+=0$ and a $N$ is sufficiently large. In Fig. (\ref{fig1}), one can see the density plots for $N=100$ (a) and $N=300$ (b). In Fig. (\ref{fig1}b), the constant $\ds{\alpha}$ is chosen to have a larger value to shift the center of the wave packet to the right. The system has only one exceptional eigenstate which is perfectly localized at the $n=0$ lattice site. However, one can see from Fig. (\ref{fig1}) that non-diffracting wave packet can still be obtained up to a large propagation distance at the exceptional point $\ds{\gamma_+=0}$. At large values of $z$, the contribution from the right edge comes into play as the lattice is not infinitely extended. Therefore non-diffracting character is lost. If we take a very large value of $N$, the wave packet would become practically stationary. Thus, we say that the continuous family of eigenstates are sensitive to the disorder in hopping amplitude disorder.
	
So far we have studied stationary eigenstates. At the exceptional point, one can also obtain a monotonically growing or decaying solutions and power oscillating solution. For example, choosing $\ds{ f(q) =e^{q^2/2}     }$ leads to growing solution while choosing $\ds{ f(q) =e^{-q^2/2}     }$ leads to decaying solution. If we choose $\ds{ f(q) =\sin{q}     }$, then power oscillation occurs, which is not expected at the exceptional point.
	
So far we have explored an interesting phenomenon at exceptional points for the NGF lattice with nonreciprocal hopping amplitude. One may ask if we can find such non-coalescing eigenstates in other semi-infinite lattice at the exceptional point? One example of such other systems is provided in \cite{CYHR20} and there are many other such systems. To generalize our strategy to find such states, we consider a system with completely random backward hopping amplitudes $\gamma_n$ at the exceptional point. We assume that forward hopping amplitudes are zero (which guarantees the existence of exceptional points since the corresponding Hamiltonian is a Jordan block). The corresponding Hamiltonian reads
		\begin{eqnarray}\label{hrw10okl1}
H=\sum_{n=1}^{N-1}\gamma_{n} |n+1><n|
\end{eqnarray}
The eigenvalue equation leads to $\ds{\gamma_{n} \psi_{n+1}=E ~\psi_n}$, where the energy $E$ is a continuous variable. One can normally solve this equation to obtain the unique eigenstate at the exceptional point. Here, our aim is to get almost non-diffracting localized waves at the exceptional point. To solve the above equation, our approach is as follows: instead of setting $\ds{\Psi_{N+1}=0}$, here we assume that $\ds{\Psi_{N+1}}$ is a very small number (for example it can be chosen to be equal to $10^{-8}$). Then we can get the field amplitude $\psi_n$ recursively and get the non-diffracting state approximately. By varying $E$, we can obtain a continuous family of such solutions. The resulting solution is not exact, but can be used as almost non-diffracting waves up to a large distance as long as  $\ds{\Psi_{N+1}}$ is a very small number and $\ds{N}$ is large.

	Let us apply our above strategy in a disordered system. We consider a disordered lattice, where $\ds{   \gamma_{n}}$ take random values in the interval $\ds{\gamma_n\in[1,6]}$. There are $N=100$ lattice sites in the system and assume that $\psi_{101}(z=0)=10^{-8}$. Then we get $\ds{\psi_n}$ for $n=1,2,...,N$ and find its time evolution numerically to check whether it shows non-diffracting behaviour. In Fig. (\ref{fig2}a), one can see the density plot as a function of propagation length $z$ at $E=2.5$. The non-diffracting wave-packet up to $z\sim 5$ can be seen from the figure, which shows that our strategy is good enough to obtain such waves even in a disordered lattice up to a large distance. In Fig.(\ref{fig2}b), we see that the total power $\ds{P=\sum_n\psi(z)}$ does not change up to $z\sim 5$. After that point, edge effects come into play and non-diffracting behavior is lost. This is already expected as our solution is not an exact solution. To this end, we say that systems with second order and third order exceptional points are not good candidates to see almost non-diffracting waves. They can only be seen in systems with higher order exceptional point as our strategy is based upon the fact that $\ds{N}$ is large.

	\section{Conclusion}
	
	In summary, we provide an analytical solution at exceptional points in Non-Hermitian Glauber-Fock lattices. It is generally believed that all eigenstates coalesce at exceptional points. We prove that a continuous family of eigenstates exists at EP for the semi
	infinite non-Hermitian Glauber Fock lattice. We show that our solution can also be used for a finite lattice. The truncated solution is in fact non-diffracting solution at exceptional points. We use the analytical solution to depict non-diffracting beams at the exceptional point in non-Hermitian photonic lattices with unidirectional couplings forming Glauber-Fock. We further provide a numerical approach to construct such non-diffraction beams in randomly and unidirectionally coupled optical lattices. 
	 
\begin{acknowledgments}
	H.R acknowledge the support by the Army Research Office Grant No. W911NF-20-1-0276 . The views and conclusions contained in this document are those of the authors and should not be interpreted as representing the official policies, either expressed or implied, of the Army Research Office or the U.S. Government. The U.S. Government is authorized to reproduce and distribute reprints for Government purposes notwithstanding any copyright notation herein. 
	
\end{acknowledgments}

\end{document}